\documentstyle[12pt]{article}

\begin{document}
%\draft
%\preprint{}

\title{ Mixmaster chaos}

\author{ {\it A. E. Motter$^a$ and P. S. Letelier$^b$}\\
Departamento de Matem\'atica Aplicada-IMECC \\
 Universidade Estadual de Campinas, Unicamp\\
 13081-970 Campinas, SP, Brazil}

\date{}
\maketitle

\begin{abstract}

The significant discussion about the possible chaotic
behavior of the mixmaster cosmological model
due to Cornish and Levin [J.N. Cornish and J.J. Levin, Phys. Rev. 
Lett. 78 (1997) 998;
Phys. Rev. D 55 (1997) 7489] is revisited.
We improve their method by correcting nontrivial
oversights that make their work inconclusive
to precisely confirm their result:
``The mixmaster universe is indeed chaotic''.

\end{abstract}

\vskip 1.0truecm



$^a$E-mail: motter@ime.unicamp.br.

$^b$E-mail: letelier@ime.unicamp.br.

\newpage

The Bianchi IX (mixmaster) cosmological model was investigated
by Belinskii, Khalatnikov  and Lifshitz (BKL), clarifying
the nature of cosmological singularities (see \cite{bkl} and Refs.
therein), and by Misner,
in his attempt to solve the cosmological
horizon problem \cite{misner}.
The subsequent emergence of chaos in maps
associated with this model
attracted the attention of several other researchers
in the following years. Later,
it was realized that standard chaotic indicators used in previous
works, like Lyapunov exponents, are not invariant under
space-{\it time} diffeomorphisms
(see \cite{hobill} for a comprehensive discussion).
Since then, a number of contributions looking
for coordinate independent manifestations of chaos have appeared
in the literature,
and the mixmaster cosmology has become one of the main
paradigms of deterministic chaos in General Relativity.
Even though some important results have been obtained,
a satisfactory comprehension of chaos in this
system remains to be achieved (see \cite{rev} for a critical review).
Among the most promising proposals we cite
the efforts of Szydlowski and collaborators in treating the problem
in terms of (invariant) curvature \cite{szy}. This approach, however, 
breaks down in the mixmaster case \cite{rev}.
Another significant progress was recently made by
Cornish and Levin (CL) \cite{cornish}.
Employing fractal techniques, the authors obtained
evidences of chaotic transients in the mixmaster model.
Their work is, however,
inconclusive due to conceptual flaws.
Accordingly, the existence of invariant chaos
in the mixmaster cosmological model is still an open question.

In dynamical systems,
chaos seems to be the rule rather than the exception
(at least for generic  autonomous systems with three or
more equations).
Based on this fact, there is small risk in declaring 
in advance that a system is chaotic.
In addition, even in nonchaotic systems,
technical errors in general
lead to some kind of numerical chaos.
Therefore,
the relevance of stating that a particular
system is chaotic stays on the precision
of the proof and the nature of the chaos
revealed by it \cite{holmes}.
In the case of mixmaster universe it stays also on
the need of a proper meaning for
invariant chaos in General Relativity. 

The aim of this communication is to show unequivocally that
the mixmaster dynamics is chaotic in a precise sense.
First we review critically the work of
CL, pointing out some key oversights.
Second, we propose modifications
and corrections of the CL's procedure.
Next, this new procedure is employed
to the approximated dynamics
and then to the exact dynamics.
From the full dynamics analysis we
conclude that the mixmaster model
does evolve chaotically.

The work of CL is based on enlarging the three outcomes of the
mixmaster universe and in studying the invariant set through fractal
methods of chaotic scattering. The existence of a fractal structure
in the invariant set would imply chaotic transients.
First they studied an approximation of the exact dynamics,
the Farey map,
whose invariant set could be exhibited explicitly. For this map,
the topological entropy and the information dimension
of the invariant set were computed.
Then, by numerically computing the information dimension
for a choice of the outcomes, the full dynamics was studied.
Based on the noninteger value
of the resulting information dimension,
the authors declared that the system is chaotic.

In checking the CL's papers we see that:
(1) Concerning the
Farey map, the invariant set exhibited,
a Cartesian product of periodic
points of the epoch-era BKL map \cite{bkl}, is not invariant.
In contrast to the authors' belief, only points with the same
period in both Cartesian components are periodic points of
the Farey map.
Consequently, the conclusions obtained from this `invariant set'
are wrong or at best suspicious. For instance, their result for the
topological entropy is twice as bigger as the correct value
because what they found was the square of the number of
periodic points
that really exist.
(2) In the full dynamics counterpart, the information dimension
was computed from the box-counting dimension of the basin
boundaries, based on a result presented in Ref. \cite{farmer}
- the information
dimension of a set may be given by
the box-counting dimension of an
{\it optimal} fraction of the set (see \cite{ott}, p. 83) - 
that does not apply to this case because the optimal condition
is not satisfied. Moreover, the initial conditions,
given near the moment of maximal
expansion, were parametrized by parameters from
Chernoff-Barrow (CB) map and the outcomes were defined
in terms of one of these parameters \cite{cb}.
Since the CB map cannot be applied
far from the cosmological singularities,
this parametrization is violated during the integration and is
recovered only when the Kasner approximation becomes valid.
Accordingly, the outcomes are not well defined all the time
and they make sense only when a Kasner approximation is made.
Consequently, the full dynamics analysis is not really full.
(3) Both approximated and full analyses involve the idea of a
{\it full repellor} as a set of the
system itself. This concept is unclear because repellors
are functions of the outcomes, and therefore are sets that
are defined only after the definition of the outcomes.
The origin of the problem seems to be in some confusion
between nonattracting invariant sets
(repellors) and the set of periodic points.

When considered the above oversights, the CL's work
becomes inconclusive.
In spite of that, the authors left
a great contribution: The suggestion of using fractal
invariant methods to investigate the mixmaster
model\footnote{Fractal techniques were also introduced
 by Demaret and  De Rop \cite{rop} to study an approximation of
  Bianchi IX cosmology.}.
In this communication
we will follow this idea as close as possible of the CL's work
in order to allow a direct comparison by the reader.
But different from CL,
we will choose the box-counting dimension as the invariant
indicator of chaos \cite{ott}.
The topological entropy and the information
dimension, employed by CL, are unsuitable indicators of chaos
in the mixmaster system:
The topological entropy is a function of the time
parametrization because of its dependence on the period of the
periodic points; the information dimension is difficult to
be computed since it is difficult to verify the optimal
condition mentioned above
(we stress that CL were not able to obtain such
dimension in the case of the full dynamics).
The box-counting dimension, on the other hand, is
diffeomorphism invariant and can be easily computed
by applying the uncertainty exponent method \cite{ott}.

The Einstein equations expressed in terms of the three
spatial scales of the model $(g_{ij})= $diag$(a^2,b^2,c^2)$
lead to the motion equations \cite{bkl}
\begin{equation}
(\ln a^2)''=(b^2-c^2)^2-a^4, \;\; et\;\; cyc.,
\end{equation}
together with the constraint equation
\begin{equation}
(\ln a^2)'(\ln b^2)'+2a^2b^2-a^4+ cyc. =0.
\end{equation}
A prime denotes $d/d\tau$, where $dt=abc\; d\tau$ and $t$
is the cosmological time.
It is convenient to introduce
$\alpha_1$, $\alpha_2$ and $\alpha_3$
as
$\alpha^{(a)}=\ln a$, $\alpha^{(b)}=\ln b$
and $\alpha^{(c)}=\ln c$
in increasing order of the time derivatives.
That is,
$\dot{\alpha}_1\leq\dot{\alpha}_2\leq\dot{\alpha}_3$,
where a dot denotes $d/d\Omega$
and $\Omega$ is the time $\ln (abc)/3$.
The dynamics near the cosmological singularity
is approximated by the CB map \cite{cb},
obtained from a parametrization of the surface of section
$\alpha_{\kappa}=0, \dot{\alpha}_{\kappa}>0$ by
\begin{equation}
\begin{array}{lll}
 \alpha_1=3\Omega q_1(u,v), & \alpha_2=3\Omega q_2(u,v),
& \alpha_3=3\Omega q_3(u,v), \\
\dot{\alpha}_1=3p_1(u), & \dot{\alpha}_2=3p_2(u),
& \dot{\alpha}_3=3p_3(u) ,
\end{array}
\label{1}
\end{equation}
where
$q_1=1/(1+v+uv)$, $q_2=\delta_{\kappa 3}(v+uv)/(1+v+uv)$,
$q_3=\delta_{\kappa 2}(v+uv)/(1+v+uv)$, $p_1=-u/(1+u+u^2)$,
$p_2=\delta_{\kappa 2}(1+u)/(1+u+u^2)+
\delta_{\kappa 3}(u+u^2)/(1+u+u^2)$
and $p_3=\delta_{\kappa 2}(u+u^2)/(1+u+u^2)+
\delta_{\kappa 3}(1+u)/(1+u+u^2)$,
for $0\leq u,v<\infty$.
The evolution of the parameters $u$ and $v$ is then determined by
the Farey map,

\begin{equation}
F(u,v)=\left\{
\begin{array}{ll}
\left( u-1, v+1\right)
& \mbox{if $u\geq 1$ (change of epoch)}\\
\left(u^{-1}-1, (v+1)^{-1} \right)
& \mbox{if $u<1$ (change of era)}.
\end{array}
\right.
\end{equation}
The scale factors\footnote{
The indices $\mu, \nu ,\rho$ are permutations of $a,b,c$.  
}
$(\alpha_1,\alpha_2,\alpha_3)=
(\alpha^{(\mu)},\alpha^{(\nu)},\alpha^{(\rho)})$
evolve as
$\dot{\alpha}_1\leq\dot{\alpha}_2\leq\dot{\alpha}_3\longmapsto
\dot{\tilde{\alpha}}_1\leq
\dot{\tilde{\alpha}}_2\leq\dot{\tilde{\alpha}}_3$,
where
$(\tilde{\alpha}_1,\tilde{\alpha}_2,\tilde{\alpha}_3)=
(\alpha^{(\nu)},\alpha^{(\mu)},\alpha^{(\rho)})$ and $\kappa=2$
except in the transition to the last epoch of each
era,
where $(\tilde{\alpha}_1,\tilde{\alpha}_2,\tilde{\alpha}_3)=
(\alpha^{(\nu)},\alpha^{(\rho)},\alpha^{(\mu)})$
and $\kappa$ becomes 3.
The Farey map can be equivalently defined as,
\begin{equation}
\bar{F}(\bar{u},\bar{v})=\left\{
\begin{array}{ll}
\left( \bar{u}-1, \bar{v}+1\right)
& \mbox{if $\bar{u}\geq 2$ (change of epoch)}\\
\left((\bar{u}-1)^{-1}, \bar{v}^{-1}+1 \right)
& \mbox{if $\bar{u}<2$ (change of era)},
\end{array}
\right.
\end{equation}
for $1\leq \bar{u},\bar{v}<\infty$.
Since $\bar{F}=f^{-1}oFof$ ($\bar{F}$ and $F$ are conjugate),
where $f(\bar{u},\bar{v})=(\bar{u}-1,\bar{v}-1)$,
the parametrization corresponding to
$\bar{F}$ is not just (\ref{1}) (as used by CL) but, (\ref{1})
with $q_i$ and $p_i$ replaced by
$\bar{q}_i=q_iof$ and $\bar{p}_i=p_iof$, for $i=1,2,3$.

First let us consider the approximated dynamics
in order to have some insight into the problem.
In taking relative quantities in (\ref{1}),
the overall scale factor $\Omega$ is canceled,
and we are led to a
stationary phase space $u\times v$ where three outcomes
can be identified: $(u,\alpha_3)=(\infty,\alpha^{(a)})$,
$(u,\alpha_3)=(\infty,\alpha^{(b)})$
and $(u,\alpha_3)=(\infty,\alpha^{(c)})$.
Since only rational values of $u$ are led to these outcomes
by forward iteration of the map,
the corresponding invariant set contains almost every point of the
phase space and it tells us nothing about chaos.
However, some useful conclusions can be obtained if we enlarge the
outcomes. It is made by defining the
three outcomes for $u>u_{out}-1$
instead of $u=\infty$,
which transform the system in an exit (scattering) system
(the original Farey exit map was defined by CL \cite{cornish}).
These outcomes are equivalent to
$\dot{\alpha}_3/(\dot{\alpha}_1+\dot{\alpha}_2
+\dot{\alpha}_3)>\hat{p}_3(u_{out})$,
where $\hat{p}_3(u)=(u+u^2)/(1+u+u^2)$.
The dynamics of this exit system is determined by a nonattracting
invariant set - the repellor \cite{ott}.
The repellor is a zero measure set consisting
of a countable number of
unstable periodic orbits surrounded by an
uncountable number of nonperiodic orbits.
A fractional box-counting dimension
for the repellor implies\footnote{
This implication is of typical validity and
includes the extreme case, when the box-counting dimension
of the repellor equals the phase space dimension.
Pathological
counterexamples can be constructed, but do not seem
to represent  realistic systems \cite{troll}.
}
the exit system to be chaotic
in the sense of having a chaotic transient: The system evolves
chaotically for a period of time before being scattered.
Concerning the original system,
the system is typically never scattered
(except for particular choices of the initial conditions)
and hence it would evolve chaotically forever.
Therefore, if the exit system is chaotic,
the original system is also chaotic.

In Fig. 1 we show, in three different colors,  the basins of
initial conditions for the outcomes defined by $u_{out}=7$
(the same used by CL \cite{cornish}). It can be shown that
magnifications of the basin boundary lead to figures with
the same complex structure of Fig. 1,
where the three colors (basins) are always present.
The basin boundary
is the future invariant set, and its box-counting dimension
estimated from the uncertainty exponent method \cite{ott}
results $D_0=1.87\pm 0.01$. Since the repellor is the
intersection of the future and the past invariant sets,
it is equivalent to study any one of these sets.
We prefer the future invariant set because of its easier
geometric interpretation. Therefore, the above
result for $D_0$ is enough to conclude that
the original map is chaotic.
In varying the outcomes, $D_0$ becomes an increasing function
of $u_{out}$ as showed in Fig. 2a. When the outcomes are
made arbitrarily small, $D_0$ approaches 2. This behavior
is completely different in a nonchaotic system, since in that
case $D_0$ is integer and
does not approach the phase space dimension
when the outcomes are reduced.

Now we repeat the procedure for the full dynamics,
numerically integrating the mixmaster equations.
In order to allow comparisons with the approximated study,
we take the initial condition defined by (\ref{1}) for
$1<u,v<2$ ($\kappa=2$), and $\Omega$ the negative solution
of the constraint equation. The integration is performed in
the negative direction of $\Omega$ (approaching the singularity).
Since $u$ is not well defined
during the integration, the best choice for the 
outcomes is
$\dot{\alpha}_3/(\dot{\alpha}_1+\dot{\alpha}_2
+\dot{\alpha}_3)>\hat{p}_3(u_{out})$.
In the limit $\Omega\rightarrow -\infty$ these outcomes
correspond to those defined for the Farey map.
The full dynamics is, however,
considered for a continuous time
and the outcomes are defined not only on a surface of section.
In Fig. 3 we show the basins of initial conditions for
$u_{out}=7$. Fig. 3 may be compared with Fig. 1.
Small differences between these figures were expected,
since the latter is an asymptotic approximation
while the former is exact and refers to initial conditions
taken near the maximum of expansion.
The complicated structure of Fig. 3 is also present in
any magnification of the basin boundary - see Fig. 4, where
we show a 100 times magnification of Fig. 3.
The numerical computation of  the box-counting dimension
corresponding to $u_{out}=7$ results $D_0=1.87\pm 0.01$,
consistent with the approximated value. In fact, the
differences between the approximated and full fractal dimensions
are quite small, as can be seen in Fig. 2.
Moreover, the system has the
so called Wada property \cite{nature}:
Every point on the boundary of a basin is also on the boundary of
the other two basins. The corresponding fractal figures of CL
do not present this property \cite{cornish}.

From the above analysis follows that the mixmaster
dynamics is in fact chaotic. This conclusion is independent
of the particular space-time parameters used
because $D_0$ is invariant under space-time diffeomorphisms.
The numerical values obtained for
$D_0$ depend on the choices made for the outcomes.
But, once found
a set of  outcomes for which the resulting $D_0$ is fractional
and goes to the phase space dimension in the limit
``outcomes $\rightarrow 0$'', the system is unequivocally
chaotic.

The authors thank R.R.D. Vilela 
and  M.T. Ujevic for their valuable comments,
and also Fapesp and CNPq for financial support.

\newpage

\begin{figure}
\caption{Portrait of the basins of the
approximated dynamics for $u_{out}=7$.
The initial conditions were chosen on a grid of $400\times 400$,
for $\alpha_1=\alpha^{(a)}$,
$\alpha_2=\alpha^{(b)}$ and $\alpha_3=\alpha^{(c)}$.
Regions in gray, black and white
correspond to orbits that escape for
$\dot{\alpha}_3/(\dot{\alpha}_1+\dot{\alpha}_2+
\dot{\alpha}_3)>56/57$
with $\alpha_3=\alpha^{(a)}$, $\alpha_3=\alpha^{(b)}$ and
$\alpha_3=\alpha^{(c)}$, respectively. 
}
\label{f1}
\end{figure}

\begin{figure}
\caption{Box-counting dimension $(D_0)$ of the basin boundary
(future invariant set) as a function of the outcome parameter
$(u_{out})$ for: (a) Approximated dynamics (triangles);
(b) Full dynamics (squares). The size of the triangles and squares
corresponds approximately to the statistical uncertainty of $D_0$.
}
\label{f2}
\end{figure}

\begin{figure}
\caption{Portrait of the basins of the full dynamics for $u_{out}=7$.
The initial conditions were chosen on a grid of $400\times 400$,
for $\alpha_1=\alpha^{(a)}$,
$\alpha_2=\alpha^{(b)}$ and $\alpha_3=\alpha^{(c)}$.
Regions in gray, black and white
correspond to orbits that escape for
$\dot{\alpha}_3/(\dot{\alpha}_1+\dot{\alpha}_2+
\dot{\alpha}_3)>56/57$ with $\alpha_3=\alpha^{(a)}$,
$\alpha_3=\alpha^{(b)}$ and $\alpha_3=\alpha^{(c)}$, respectively. 
}
\label{f3}
\end{figure}

\begin{figure}
\caption{A 100 times magnification of a portion of Fig. 3.}
\label{f4}
\end{figure}

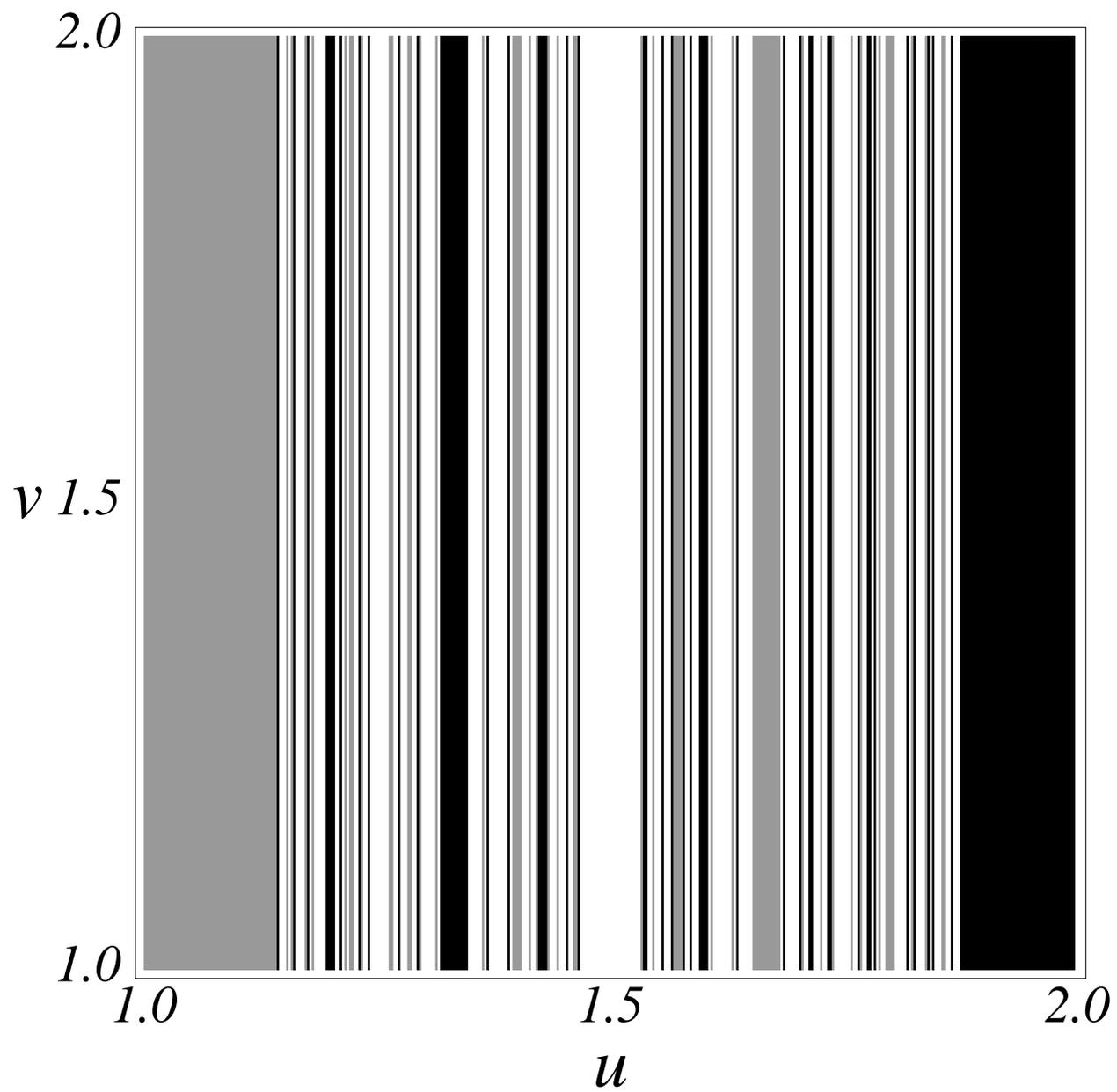
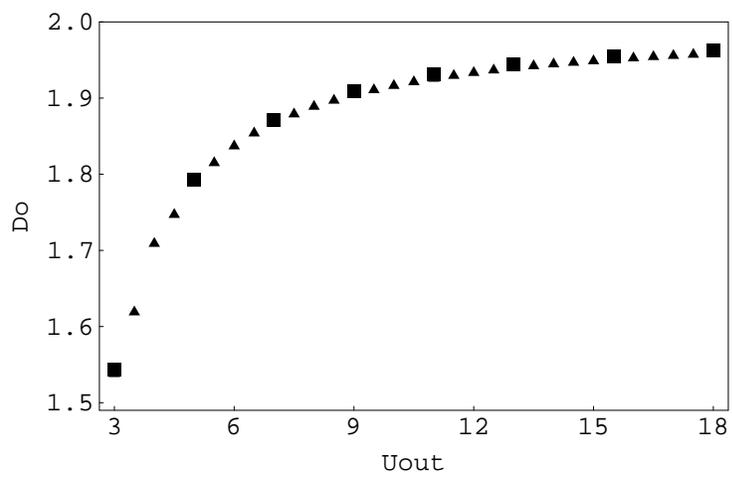
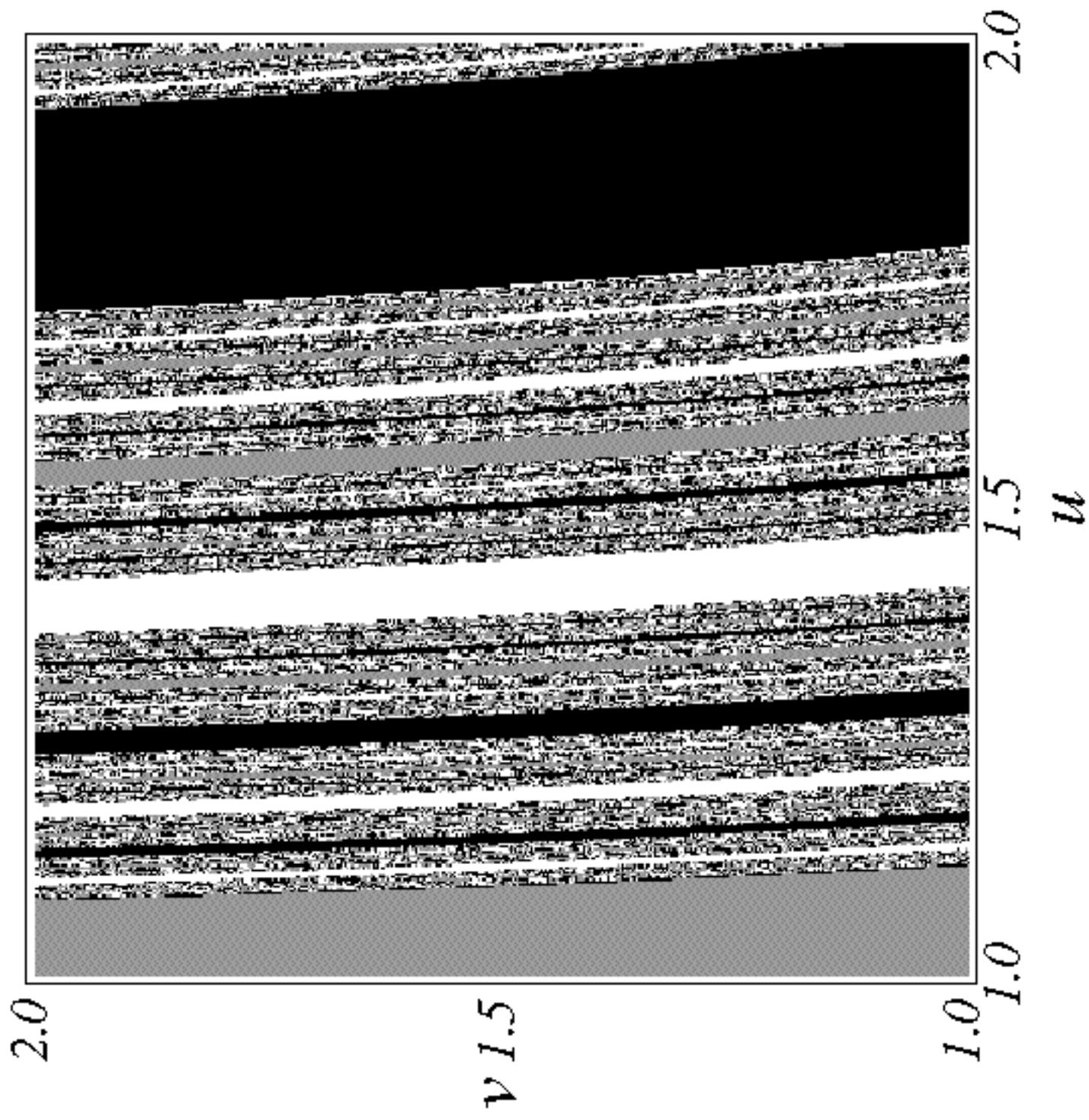
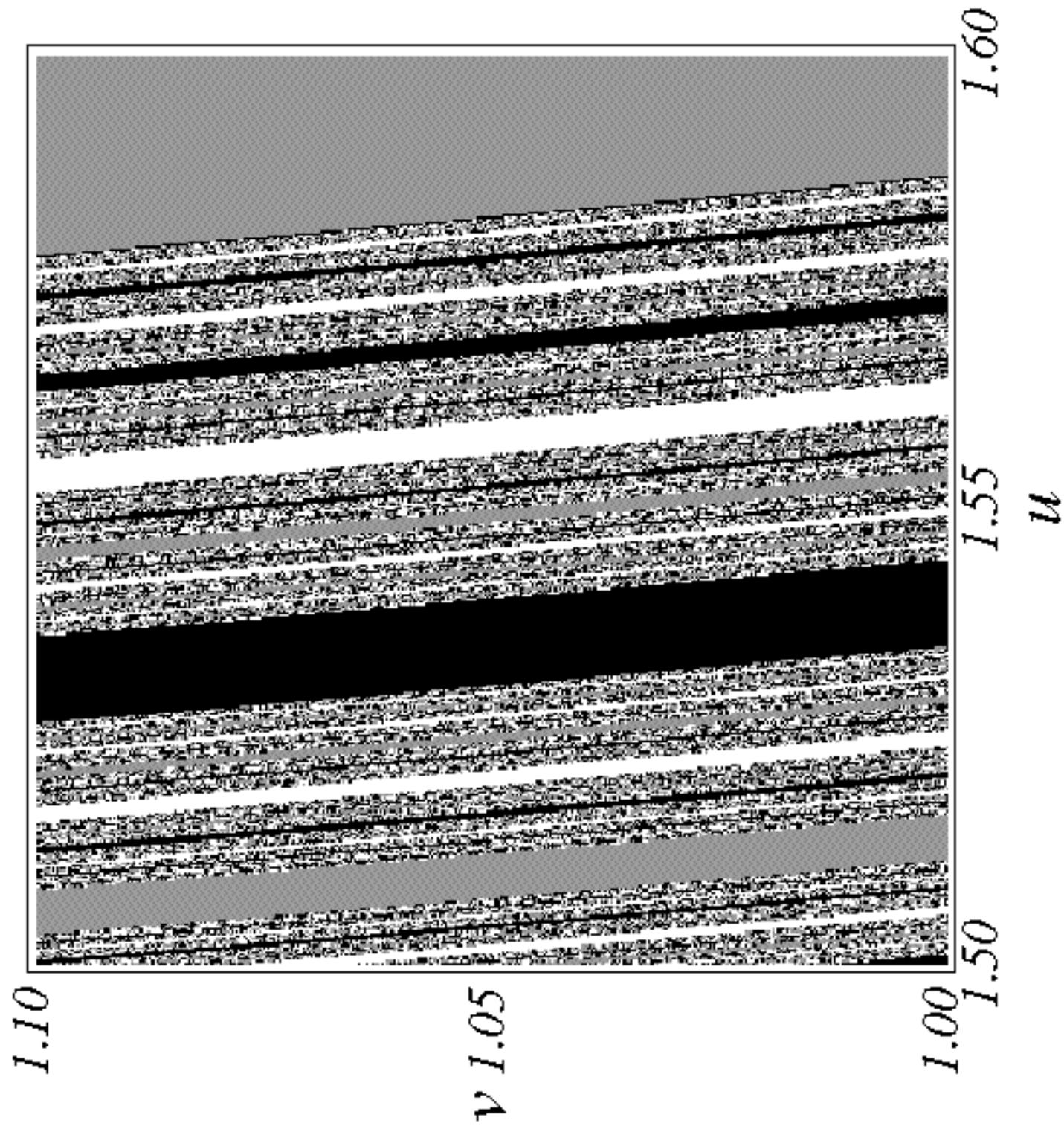
\end{document}